\documentclass{appolb}
\usepackage{graphicx}

\begin{document}
\title{Following in Tini's Giant Footsteps%
\thanks{Contribution to the special volume of Acta Physica Polonica B commemorating \\ Martinus Veltman}%
}
\author{John Ellis
\address{King's College London, Strand, London, WC2R 2LS, United Kingdom; \\
Theoretical Physics Department, CERN, Geneva, Switzerland;\\
National Institute of Chemical Physics \& Biophysics, R{\" a}vala 10, 10143 Tallinn, Estonia}
\\
}
\maketitle
\begin{abstract}
This paper describes my personal appreciation of some of Tini Veltman's great research achievements and how my
own research career has followed the pathways he opened. Among the topics where he has been the most 
influential have been the pursuit and study of the Higgs boson and the calculation of radiative
corrections that enabled the masses of the top quark and the Higgs boson to be predicted ahead of
their discoveries. The search for physics beyond the Standard Model may require a complementary
approach, such as the search for non-renormalizable interactions via the Standard Model Effective Field Theory. \\
\begin{center}
KCL-PH-TH/2021-27, CERN-TH-2021-067
\end{center}
\end{abstract}
  
\section{Introduction}
\label{sec:introduction}

It is an honour to have been offered the opportunity to contribute to this memorial volume
celebrating Martinus (Tini, as he was generally known) Veltman, particularly because he was 
one of my scientific heroes. In addition to his personal scientific research, he played a
central role in putting Dutch theoretical particle physics on the international map
(think Gerard 't Hooft, Bernard De Wit, Peter van Nieuwenhuizen and many others),
pioneered the development of computer algebra with his {\tt Schoonschip} programme,
and was a key early supporter of the LEP accelerator.
Moreover, beneath his occasionally idiosyncratic and outspoken exterior lurked a fiercely
independent thinker with empathy for young theorists whose work he appreciated, as well as
for the experimentalists whose work he encouraged.

\section{Renormalization}
\label{sec:Renormalization}

Tini's main scientific achievement was of course the demonstration (together with 't Hooft)
how spontaneously-broken non-Abelian gauge theories could be renormalized~\cite{Veltman1968,Veltman1970,tHooft1971,tHooft1972}, 
and my main scientific contacts with Tini resulted from my own research work pursuing
corollaries of that work. It is worth remembering that Tini's most consequential work
was done somewhat in the wilderness, struggling tirelessly with a problem that most 
theorists disregarded: ``sweeping an odd corner of weak interactions" as he reported being
told by Sidney Coleman~\cite{SciAm}. For one thing, quantum field theory in general was unfashionable
for much of the 1960s, as causality, analyticity and the S-matrix held sway. For another,
very few theorists could see the interest in non-Abelian gauge theories. Significant progress had
been made by Feynman, DeWitt, Faddeev and Popov in formulating massless non-Abelian theories
at the quantum level, but nobody understood how to renormalize them if the gauge bosons were
massive, and obtain sensible, finite results. 

Many suggested that massive gauge vector bosons were the likely mediators of the weak interactions, 
once the $V-A$ structure of their effective four-fermion interaction was established experimentally.
Among the proposals were Glashow's proposal in 1961~\cite{Glashow} of the SU(2)$\times$U(1) structure of the electroweak sector of
the Standard Model, in which he simply postulated masses for such bosons without
concerning himself with their origins, nor the renormalizability of his model. Some years
later, in 1967, Salam~\cite{Salam} and Weinberg~\cite{Weinberg} revived his model with the suggestion that their masses might arise
from spontaneous symmetry breaking, using the mechanism that had been proposed back in 1964 by Higgs~\cite{Higgs}, Englert and
Brout~\cite{EB}, and Kibble~\cite{Kibble}. They did not attempt to prove that the Standard Model was
renormalizable, though apparently Weinberg suspected that it might be, and set a student to
trying (unsuccessfully) to prove it. The numbers of citations of these papers were minuscule
until the papers of Tini and 't Hooft showed how to renormalize and obtain finite
results from spontaneously-broken gauge theories.

\section{The Higgs Boson}
\label{sec:Higgs}

These papers triggered a tsunami of theoretical interest, joined by
experimental interest following the discovery of neutral currents in
1973~\cite{NC} and charmonium in 1974~\cite{Jpsi}. Mainstream efforts started targeting
the discovery of the massive electroweak gauge bosons, but that was
not the priority for Mary Gaillard, Dimitri Nanopoulos and myself.
We reasoned that the key element in the whole theoretical edifice
of the Standard Model was the Higgs-Englert-Brout mechanism for
spontaneous gauge symmetry breaking, and the existence of the
physical scalar boson predicted by Higgs~\cite{Higgs} (who had also considered
several of its physical properties~\cite{Higgs3}). This was why we set out to
write our paper on the phenomenological profile of the Higgs boson~\cite{EGNH}.
Because we were aware that we were out on what was considered by most senior theorists in those days to be quite a hypothetical limb,
we ended our paper by saying (tongues somewhat in cheek) that
``we [did] not want to encourage big experimental searches for the Higgs boson".

However, Tini was supportive of our efforts. I remember giving a
talk on Higgs phenomenology
at a gauge theory workshop at the {\' E}cole Normale in Paris in early 1976 about our
paper. The reception was generally tepid, but Tini was very
positive about it. There had been only a handful of papers before
ours on signatures of the Higgs boson, and nobody had a clue what
its mass might be. Tini had already worked on this problem~\cite{PRL75},
arguing that a very small Higgs mass would be incompatible with cosmology,
and this problem continued to be one of his principal theoretical interests in the
following years~\cite{SciAm}.
In particular, he was one of the first to point out that a massive Higgs boson
would necessarily be strongly-interacting, and to argue that this
imposed a qualitative upper limit on its mass of a few hundred GeV~\cite{logarithmic},
at which stage it would become strongly-interacting and some low-mass bound states might emerge.
~\footnote{The upper limit on the Higgs mass was also discussed in~\cite{Vayonakis,LQT}.}

The renormalizability of the Standard Model is a joint effort of
all its particles: remove any of them and uncontrollable infinities appear.
This is reflected in the growths of their quantum loop contributions to physical
quantities as their masses increase, which may be either quadratic
or logarithmic. Tini was a pioneer in exploring these renormalization effects,
and how they would affect physical observables. He considered these questions
in~\cite{logarithmic}, showing that quadratic dependences on heavy-particle masses
are the general rule, but the one-loop effects of the Higgs boson
in the Standard Model increase only logarithmically with its mass,
because of the screening effects of a custodial symmetry in the Higgs sector (see~\cite{Screening1994} for a later discussion).
Tini went on to discuss the quadratic mass dependences of one-loop corrections to 
vector boson masses in~\cite{NPB123} imposed an upper limit of several hundred
GeV on mass differences within a fermion multiplet.

These pioneering papers were followed by a stream of calculations of one-loop
corrections to various specific electroweak processes, paying particular attention
to their sensitivities to the Higgs mass. These included the
violation of $\mu - e$ universality in lepton-hadron interactions~\cite{PLB70}
(very topical at the moment!~\cite{LHCb,g-2}); the processes $e^+ e^- \to \mu^+ \mu^-$~\cite{NPB160} (with Giampiero Passarino)
and $e^+ e^- \to W^+ W^-$~\cite{NPB164} (with M.~Lemoine); vector-boson masses~\cite{PLB91};
and low-energy processes~\cite{NPB169} (with Martin Green). Somewhat later, Tini made a
heroic calculation of two-loop corrections to the ratio of $W$ and $Z$ masses
(the $\rho$ parameter)~\cite{NPB231} (with Jochum van de Bij), showing that they are quadratically sensitive to the 
Higgs mass. On this basis, Tini and Jochum argued that perturbation theory would breakdown for a Higgs mass
$> 3$~TeV. These pioneering
calculations played key roles in the subsequent predictions of 
the top and Higgs masses on the basis of high-precision LEP data,
as discussed below.

Tini was also one of first to worry about the quadratic divergences
in the quantum corrections to the Higgs mass, considering the
possibility of cancellations between fermion and boson loops~\cite{cancel}.
In particular, he considered the possibility of cancelling the top quark
contribution with those of the massive vector and Higgs bosons
in the Standard Model, and also mentioned the possibility of supersymmetry.
However, he was never strongly enamoured of it, 
although it is the most systematic realization of his idea. (It remains
to be seen whether Nature likes it!)

In parallel to his theoretical work during this period, Tini was a member of the CERN
Scientific Policy Committee, where he was an enthusiastic advocate of
the construction of LEP. This project grew out of a paper written by
Burt Richter while he has on sabbatical at CERN in the academic year
1975/6~\cite{Burt}, in which he analyzed the possible scaling up of circular
$e^+ e^-$ colliders from the few GeV of those operating at the time
to a machine with beams of energy $\sim 100$~GeV each. Mary Gaillard
and I were tasked with writing theoretical section of the first LEP physics study,
which was published in 1976~\cite{EG}. In addition to precise experiments at the 
$Z$ peak and measurements of $W^+ W^-$ production, we highlighted the
importance of searching for the Higgs boson, e.g., via its production
in association with a $Z$ boson. Tini supported strongly this physics
programme, and pushed for LEP to have the largest size compatible
with CERN's cramped geographical surroundings, so as to reach the
highest centre-of-mass energy possible. This foresight also maximized
the real estate available for the later construction of the LHC,
where the Higgs boson was finally discovered. 

\section{Radiative Corrections}
\label{sec:Corrections}

During the 1980s, there was a sustained theoretical campaign to make and refine
calculations of quantum corrections to many physical quantities that 
were to be measured at LEP. These made manifest the top and Higgs mass
dependences that had been foreshadowed by Tini in his pioneering calculations.
These theoretical contributions were gathered and reviewed in a series
of CERN reports~\cite{86-02,87-08,89-08,96-01}, where they were presented in an experimentalist-friendly
way.

Meanwhile, many low-energy experiments were providing a growing array of
constraints on the electroweak sector of the Standard Model, particularly
on the neutral-current interactions. Towards the end of the 1980s, they
were sufficiently precise to be sensitive to quantum loop corrections,
and in 1987 it became possible~\cite{Amaldi1987,Costa1987} to constrain the top quark mass through the
quadratic quantum effects that Tini had pointed out. We found an upper limit on its
mass similar to the value that was subsequently measured experimentally~\cite{EF1988}.
We also pointed out that it would also be possible to establish a lower
bound on $m_t$ once the $Z$ mass was measured accurately~\cite{EF1989}, which was done by
the CDF experiment at Fermilab, the SLC and LEP in 1989. However, these data
were not yet accurate enough to provide any useful information about the mass
of the Higgs boson~\cite{EF1990}, whose effects were suppressed by Tini's screening theorem.

During the following few years, LEP (and SLC) produced many more high-precision
electroweak measurements, and the net around the top quark mass drew tighter,
enabling it to be estimated with $\sim 10 \%$ accuracy~\cite{LEPEWK}. The plethora of LEP
measurements also provided the first indications on the possible mass of the Higgs
boson, indicating that it probably weighed $< 300$~GeV~\cite{EFL1993}.
The first direct evidence for the top quark quark came in 1994~\cite{tEvidence}, and in 1995 it became
strong enough to claim discovery~\cite{tDiscovery}. The measured mass was quite consistent with the
indirect estimate based on Tini's loop corrections. Moreover, the combination
of the direct measurement with the high-precision electroweak data made it possible
to refine the estimate of the Higgs mass~\cite{EFL1996}.

The Nobel Physics Prize was awarded to Tini and Gerard 't Hooft in 1999. To quote
the Nobel citation: ``They showed that the non-Abelian quantum field theories could 
make sense and provided a method for computing quantum corrections in these theories
... the mass of the top quark could be predicted, using high precision data from the 
accelerator LEP ... several years before it was discovered".

The story did not end there. The direct search for the Higgs boson had evolved from a minority
interest~\cite{EGNH}, supported enthusiastically by Tini, into a central theme of the LEP
experimental programme~\cite{EG,89-08,96-01}. However, searches at LEP were ultimately unsuccessful,
constraining its mass to be $> 114$~GeV~\cite{HLEP}. The torch was then passed to Fermilab,
where unsuccessful searches excluded a range of masses around 160~GeV~\cite{HTeV}. A global analysis of the
combined direct and indirect information of the mass of the Higgs boson in 2011
quoted the 68\% mass range $m_H = 120^{+12}_{-5}$~GeV~\cite{Gfitter}. Finally, in 2012 the Higgs boson
was discovered, with a mass of 125 GeV and an uncertainty $< 200$~MeV~\cite{Hdiscovery}. This provided
the final experimental vindication of Tini's proof of the renormalizability of
spontaneously-broken gauge theory, his obsession with the Higgs boson,
and his calculations of quantum loop corrections.

\section{What Next?}
\label{sec:Next}

And the story continues. On the one hand, high-energy measurements at the LHC continue
obstinately to agree with the Standard Model, in particular those of the production
mechanisms and decays of the Higgs boson. On the other hand, every term in its 
effective Lagrangian:
\begin{equation}
    {\cal L} \; \ni \; y H {\bar \psi} \psi + \mu^2 |H|^2 - \lambda |H|^4 - V_0 + \dots
\end{equation}
poses a theoretical mystery. The pattern of Yukawa couplings $y$ is the flavour problem
of the Standard Model, to which recent LHCb measurements~\cite{LHCb} have added. While the
Higgs can be responsible for the masses and mixings of fundamental matter fermions,
it does not explain their magnitudes. The magnitude of the mass term $\mu$ raises
the notorious naturalness/hierarchy problem: why is it not of the same order of
magnitude as the quantum loop corrections that Tini strove to cancel? The magnitude
of the quartic coupling $\lambda$ corresponding to the measured values of the Higgs
vacuum expectation value (vev) and $m_h$ is so small that it is probably driven
negative at high renormalization scales by radiative corrections due to the top 
quark~\cite{VacInstab}, in which case the present electroweak vacuum would not be stable, and 
how the Higgs evolved to reach its observed vev would pose a cosmological puzzle~\cite{Malc}.
The constant term $V_0$ corresponds to the comsological constant (a.k.a. dark energy).
It is measured to be
${\cal O}(\rm meV)^3$, which is far smaller, e.g., than the difference in the 
vacuum energy between the local maximum of the Higgs potential at the origin and 
its present-day value after sinking into its vev~\cite{Linde}, or the shift induced by
non-perturbative QCD effects.~\footnote{I remember having inclonclusive
discussions about the magnitude of the cosmological constant with Tini in
early 1976. To my mind the problem is not that it exists, but that it is
so small.} Finally, there could be additional terms represented
by the $\dots$, of higher order in the Higgs and other Standard Model fields.
These would not be renormalizable in the same sense as the Standard Model, as
proved by Tini and 't Hooft, but they could appear as terms in a low-energy 
effective field theory when massive particles in some renormalizable extension 
of the Standard Model are integrated out.

The possibilities for such terms are taken into account systematically by the
Standard Model Effective Field Theory (SMEFT)~\cite{SMEFT}, which includes all operators of dimensions
$d \ge 5$ that are invariant under the Standard Model gauge group and composed of
Standard Model fields with their conventional quantum numbers:
\begin{equation}
    {\cal L}_{\rm SMEFT} \; = \; \sum_{i: d \ge 5} \frac{C_i}{\Lambda_i^{d-4}} {\cal O}_i \, ,
\end{equation}
where the $\Lambda_i$ are mass scales characteristic of the new physics generating the
operators ${\cal O}_i$ with coefficients $C_i$. The operators of dimension 5 in the SMEFT
can generate neutrino masses, but are not of interest to us here. There are in general
2499 operators of dimension 6, many of which could in principle contribute to cross
sections measured at the LHC and are constrained by LHC measurements, as well as
measurements of Higgs decays and high-precision electroweak measurements at LEP and
elsewhere. If we could discover SMEFT interactions with dimension $\ge 6$ and disentangle their structure,
we would have clues to physics beyond the Standard Model, just as the $V - A$ structure of
the weak interactions led us towards Tini's giant footsteps.

\begin{figure}[t] 
\vspace{-5mm}
\centering
\includegraphics[width=0.9\textwidth]{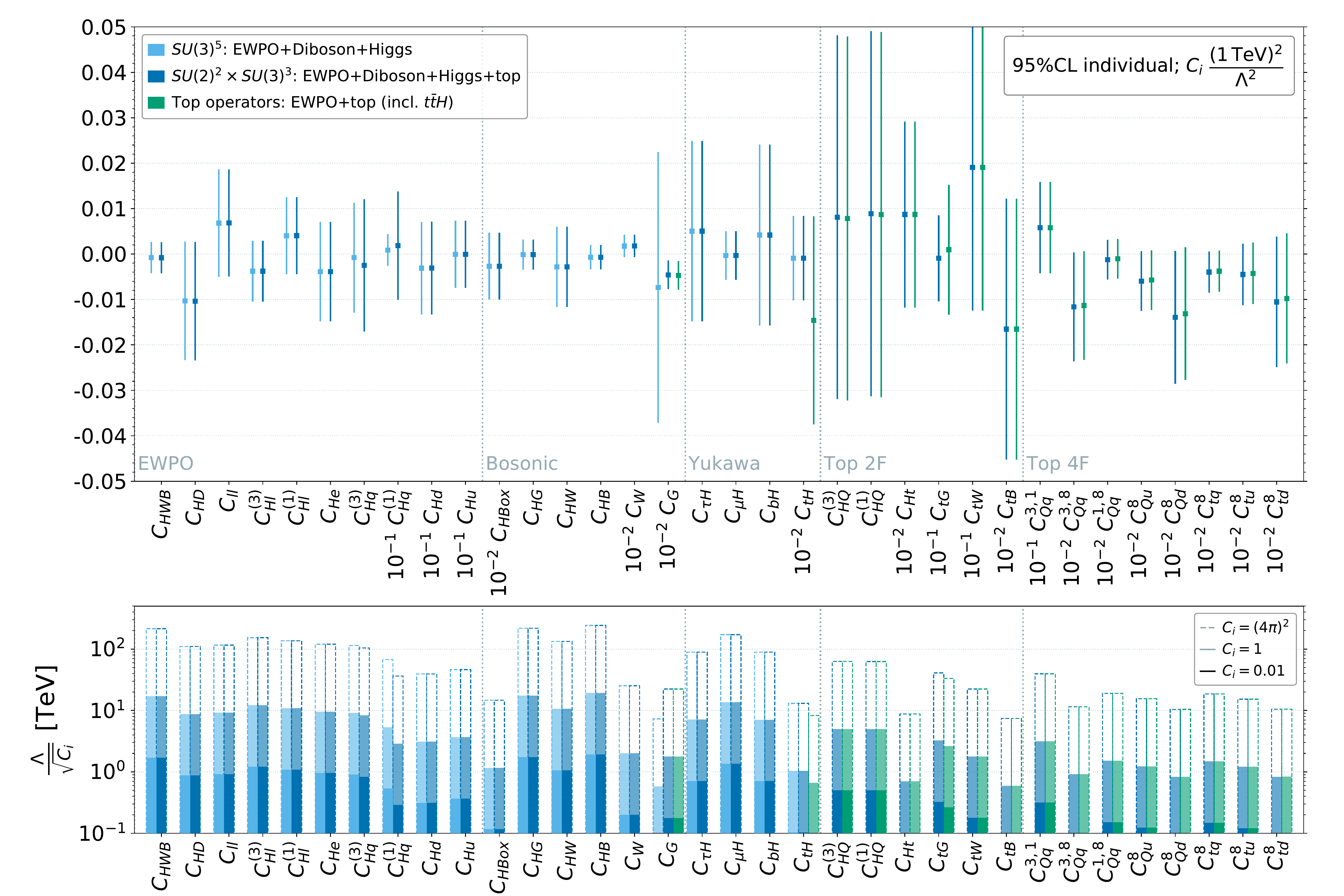}\\
\caption{\it \label{fig:all_fit} Results from a global fit~\cite{EMMSY}
to the electroweak, diboson, Higgs and top data in a top-specific scenario
with SU(2)$^2\times$SU(3)$^3$ symmetry~\cite{EMMSY}. The two panels show results for fits to 34
individual dimension-6 operators, showing the 95\% CL ranges for the operator
coefficients $C_i$ normalising all the new physics scales $\Lambda_i$ to
1~TeV, and the ranges for the scales $\Lambda_i$ for different universal values of the $C_i$.}
\end{figure}

We have recently performed a global analysis of the constraints on dimension-6 operators 
provided by all the available data, making simplifying assumptions about their flavour structure,
specifically that operators including fermions have flavour-universal coefficients, apart
possibly from those involving the top quark~\cite{EMMSY}. Switching on individually the 34 dimension-6 operators
in a top-specific scenario with SU(2)$^2\times$SU(3)$^3$ symmetry, we see in Fig.~\ref{fig:all_fit}
that the operator scales range between a few hundred GeV and $\sim 20$~TeV if their coefficients
$C_i = 1$.

\begin{figure}[t!] 
\centering
\includegraphics[width=0.8\textwidth]{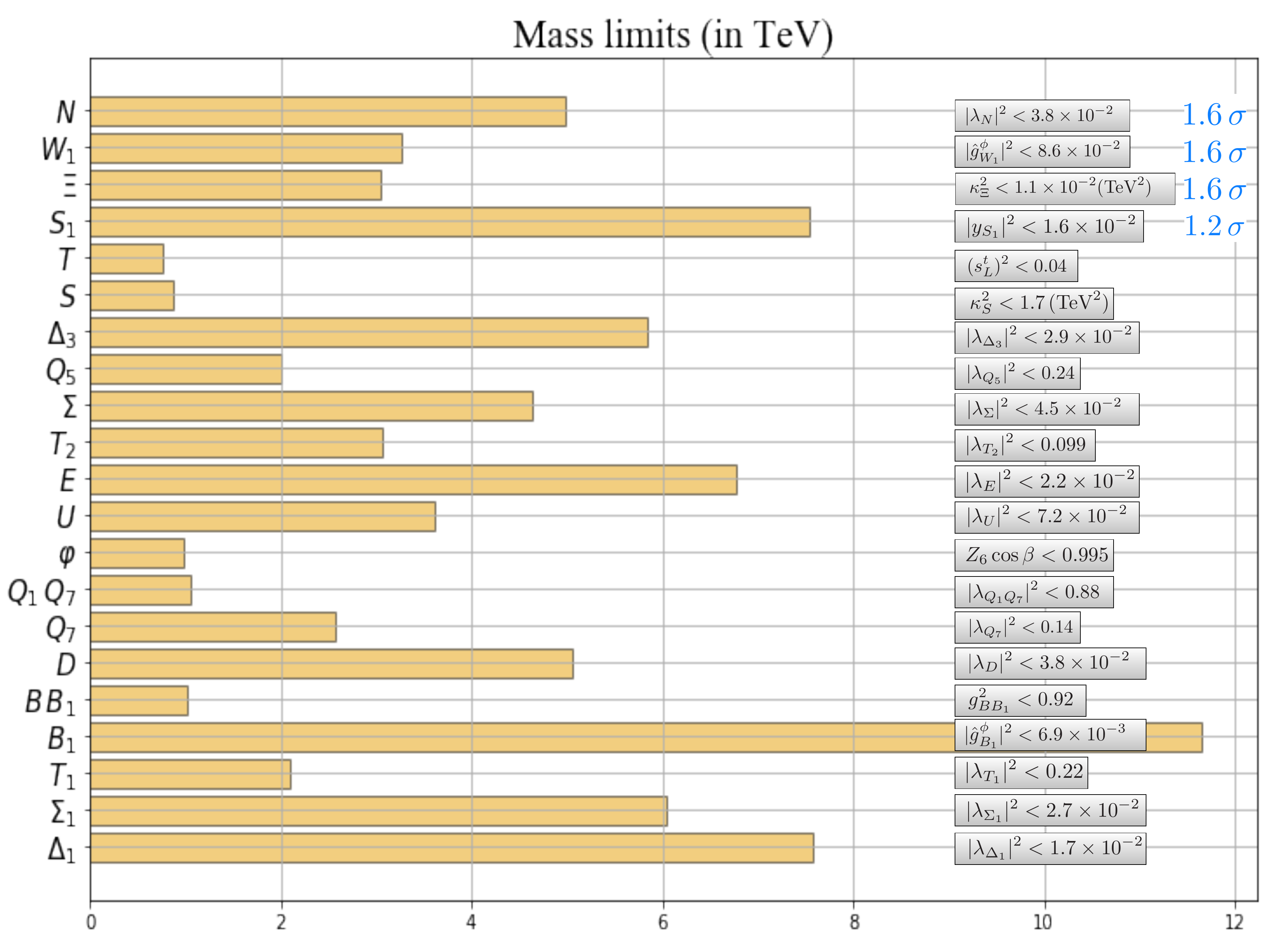}
\caption{\it The yellow bars show the mass limits (in TeV) at the 95\% CL for the
single-field extensions of the Standard Model catalogued in~\cite{EMMSY},
setting the corresponding couplings to unity. The grey boxes contain the
coupling limits obtained when setting the mass to 1 TeV, and the numbers in light blue 
are the pulls in the fit that exceed 1-$\sigma$.}
\label{fig:1Dlimits}
\end{figure}

Searching for possible indications of new physics beyond the Standard Model, we have
considered all the single-field extensions of the Standard Model catalogued in~\cite{dictionary},
and the corresponding mass limits (in TeV) at the 95\% CL and upper limits on couplings
assuming masses of 1 TeV are shown in Fig.~\ref{fig:1Dlimits}. As shown also, there are no
significant pulls in the fit that might indicate the presence of some new physics.

\begin{figure}[t!] 
\centering
\includegraphics[width=0.6\textwidth]{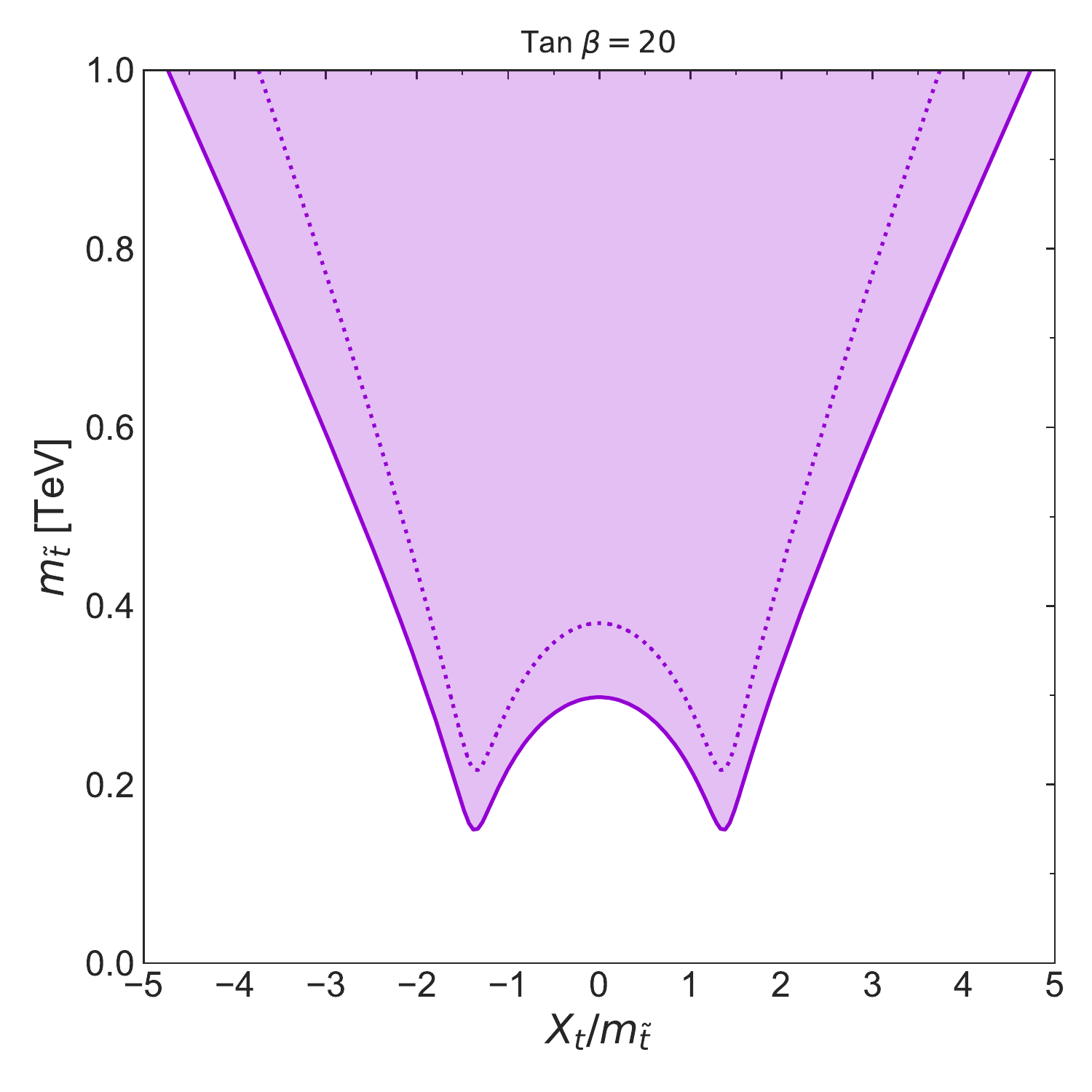}
\caption{\it Limits found in the global fit~\cite{EMMSY} on the mass of the lighter stop squark, $m_{\tilde{t}}$,
and the stop mixing parameter, $\frac{X_{t}}{m_{\tilde{t}}}$, for one value of the ratio of supersymmetric
Higgs vacuum expectation values, $\tan \beta = 20$.}
\label{fig:stop}
\end{figure}

These results indicate that there is a non-trivial hierarchy of mass scales between
the particles of the Standard Model and whatever might appear at higher energies.
As mentioned earlier, Tini was concerned about the quadratic divergence in the
mass of the Higgs boson that appears in the Standard, and speculated how to cancel it.
One possibility that has attracted much attention over the years is supersymmetry (though Tini did
not pursue it). Since the Standard Model is renormalizable by itself, and supersymmetry
is an add-on, its contributions to low-energy measurements via loop diagrams do not
grow with the supersymmetry mass scale, and are difficult to constrain via precision
measurements. Indeed, global analyses of LEP and LHC data using either exact
calculations of supersymmetric radiative corrections~\cite{pMSSM11} or the SMEFT approach are quite consistent
with the lightest supersymmetric particles, such as the lighter stop squark, weighing just a few hundred GeV~\cite{EMMSY},
as seen in Fig.~\ref{fig:stop}. 

\begin{figure}[t!] 
\centering
\includegraphics[width=0.9\textwidth]{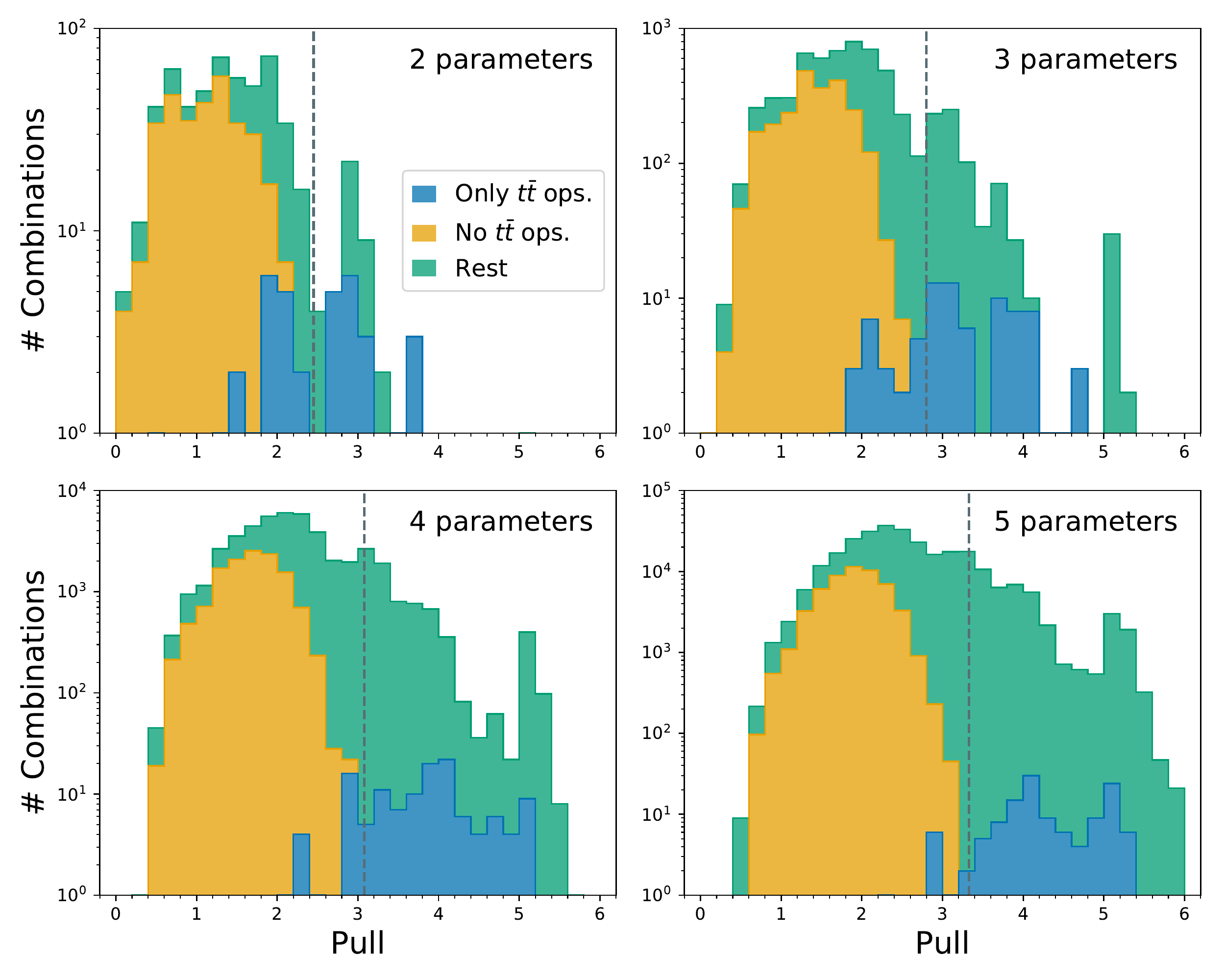}
\caption{\label{fig:pull_dist}
\it The distributions of pulls found~\cite{EMMSY} in global fits to 2 (upper left), 
3 (upper right), 4 (lower left) and 5 (lower right) parameter subsets
of operators, which have been split into three categories: those including 
only operators that affect $t\bar{t}$ production (blue), those without any 
operators that affect $t\bar{t}$ production (orange), and the rest (green). 
The dashed vertical lines mark the ranges for the pull distributions
expected at the 95\% confidence level.
}
\end{figure}

In general, supersymmetric particles may contribute to several SMEFT coefficients, 
and the same holds for many other possible extensions of the Standard Model.
Accordingly, we performed~\cite{EMMSY} a general survey of all possible extensions that
contribute to 2, 3, 4 or 5 SMEFT coefficients, as seen in Fig.~\ref{fig:pull_dist}, which shows the distributions of pulls found in three categories of SMEFT operators:
those that include only operators that affect $t\bar{t}$ production (blue), 
those that do not include operators that affect $t\bar{t}$ production (orange), 
and the rest (green). So far, we see no significant evidence for possible physics beyond
the Standard Model.

\section{Reflections}
\label{sec:Reflections}

What would Tini make of the current situation in particle physics? On the
one hand, he would be justifiably proud of the robust successes of the
Standard Model that did so much to place on a firm footing. On the other
hand, some of the problems that concerned him remain unresolved, notably
the quadratic instability in the mass of the Higgs boson and the
magnitude of the dark energy (cosmological constant). What would he think of
the experimental anomalies involving muons that have been strengthened recently~\cite{LHCb,g-2}. My
suspicion is that he would not yet be convinced, if only because they do
not have any obvious bearing on his theoretical preoccupations. My hunch
is that Tini would continue to focus his energies on the manifold problems
associated with the Higgs boson that he did so much to promote to
theoretical respectability and experimental reality.

\section*{Acknowledgements}
This work was 
supported in part by STFC grant ST/T000759/1 and by Estonian Research Council grant MOBTT5.

\end{document}